\documentclass{elsarticle}

\usepackage{lineno,hyperref}
\usepackage{wrapfig}
\usepackage{setspace}

\journal{Artificial Intelligence in Medicine}

\usepackage{mdframed}

\begin{document}
	
	\begin{frontmatter}
		
		\title{An architecture of open-source tools to combine textual information extraction, faceted search and information visualisation}
		
		\author{Daniel Sonntag\corref{mycorrespondingauthor}}
		\cortext[mycorrespondingauthor]{Corresponding author}
		\ead{sonntag@dfki.de}
		
	    \author{Hans-J\"{u}rgen Profitlich}
		\address{German Research Center for Artificial Intelligence\\
			DFKI\\
			66123 Saarbr\"{u}cken, Germany}
			
		\begin{abstract}				
			This article presents our steps to integrate complex and partly unstructured medical data into a clinical research database with subsequent decision support. Our main application is an integrated faceted search tool, accompanied by the visualisation of results of automatic information extraction from textual documents. We describe the details of our technical architecture (open-source tools), to be replicated at other universities, research institutes, or hospitals. Our exemplary use cases are nephrology and mammography. The software was first developed in the nephrology domain and then adapted to the mammography use case. We report on these case studies, illustrating how the application can be used by a clinician and which questions can be answered. We show that our architecture and the employed software modules are suitable for both areas of application with a limited amount of adaptations. For example, in nephrology we try to answer questions about the temporal characteristics of event sequences to gain significant insight from the data for cohort selection. We present a versatile time-line tool that enables the user to explore relations between a multitude of diagnosis and laboratory values. 
		\end{abstract}
		
	\end{frontmatter}


\footnotetext{DFKI Technical Report, DOI: 10.1016/j.artmed.2018.08.003. \\
\textcircled{c} 2018. This manuscript version is made available under the CC-BY-NC-ND 4.0 license http://creativecommons.org/licenses/by-nc-nd/4.0/}

\newcommand{\charite}{Charit\'{e}}
\newcommand{\tbase}{TBase\textsuperscript\textregistered}

\section{Introduction}

As medical records may cover a very long history of diseases (in this work, we have access to individual patient data up to 30 years) and include a vast number of diagnoses, symptoms, results, medications, and laboratory values, we could highly benefit from advanced search capabilities in clinical information systems to allow for the retrieval of relevant data. 
A three stage process has been proposed in \cite{SchmidtPS16}: (1) off-line textual information extraction from medical records in transplant medicine; (2) the application of interesting faceted search capabilities to the results of the previous stage; (3) the combination of the information extraction results with structured database facts. 
In the current contribution, we present a system architecture which combines (1) textual information extraction, (2) faceted search over information extraction results and structured medical data, and (3) information visualisation. 
We focus on the question whether this architecture is suitable for two distinct domains.  
In addition to the exploration of a collection of diagnoses, symptoms, results, medications, and laboratory values by applying multiple filters, we aim at supporting  the doctors' cognitive chain of decision-making by visualising key patient data. In particular, automatic information extraction from text, followed by faceted search applications, followed by the visualisation  of a combination of the information extraction results with structured laboratory values allow physicians to identify groups of patients with similar and different attributes that are highly relevant for further treatment decisions \cite{DBLP:conf/cbms/SonntagP17}.

We describe two applications of our system, their common architecture and the adaptations in functionalities and usage that had to be made to adopt the systems to the different domains and scenarios. 


\section{Terminology and Related Work}
While previous research has focused on searching and browsing, and a special form of faceted search, overview tasks that are based on automatically extracted information from text in combination with laboratory values are often overlooked.

\subsection{Information Extraction} 
Information extraction is the process of automatically deriving high-quality structured information from text. A range of applications has been described in the medical application area, for example for extracting adverse drug events from text \cite{ad15} or for symptom extraction from texts on rare diseases \cite{sym15}. However, clinical information extraction from patient records is still under-represented and underdeveloped. Earlier work includes evaluating context features for medical relation mining on scientific abstracts \cite{sonntag03}. The identification of semantic relations, such as substance A treats disease B, remains a non-trivial task \cite{sonntag03}. Recent work and comparative baseline experiments include temporal information extraction \cite{MS14}. A special trend becomes apparent, the need for ontology modelling of medical terminology and corresponding information extraction results \cite{SonntagWBZ09}. 
Large-scale academia/industry projects show this demand (THESEUS MEDICO from BMWi):  The objective is to enable a seamless integration of medical images and different user applications by providing direct access to image semantics. Semantic image retrieval should provide the basis for the help in clinical decision support and computer aided diagnosis. 
Because of enormous annotation costs, mainly unsupervised methods are being used \cite{Alicante2016}. In industry and in the context of reliable clinical relevance, however, very detailed (and labor-intensive) manual hard-coded rule-based approaches represent the state-of-the-art. In the nephrology use case, we use our research project partner's solution (Averbis), which is based on shallow text parsing, see averbis.com/en/research. In the mammography use case, we base our system on \cite{DBLP:journals/nle/KlueglTBFP16} for the Breast imaging-reporting and data system (BIRADS) and medications, with additional highly specific rule engines for relation extraction \cite{DBLP:conf/cbms/BretschneiderZH17}. 

\subsection{Faceted Search}
Faceted search allows the user to explore a data collection by applying filters in an arbitrary order. The information elements are organised by a faceted classification system among multiple dimensions, called facets. In a faceted search interface, labels are assigned to hierarchical items from the collection. This representation known as hierarchical faceted search is gaining great attraction within several information retrieval communities \cite{Yee:2003, Ben-Yitzhak:2008}. Some previous approaches focus on automating the creation of such hierarchical faceted metadata structures \cite{Stoica2007Automating}.    
We however implement a special multi-facet functionality for our medical data where we already know the hierarchy of the terms according to clinical ontologies. Design recommendations for hierarchical faceted search interfaces \cite{HearstFacettedBrowsing} include the support for flexible navigation, seamless integration with keyword search on the full text, fluid alternation between refining and expanding, avoidance of empty result sets, and at all times retaining a feeling of control and understanding (of the patient data). 

Our approach shows the following main advantages over previous approaches in the medical domain:
\begin{itemize}
 \item In our faceted search application, the user may remove \emph{any} restriction he or she may have made in previous steps, for example by choosing concrete facet values or narrowing value ranges. This allows for a much better navigation through the search space while related systems only support subsequent thinning \cite{sacco2006, sacco2015}.
 \item We base automatically generated facets (e.g., disease/symptom relationships and negations) on multi-term extraction and relation extraction, by employing state-of-the-art, high-precision textual information extraction modules.  
\end{itemize} 

\subsection{Information Visualisation} Information visualisation is the study of interactive visual representations of abstract data to reinforce human cognition. The abstract data, here the facets,  include both numerical and non-numerical text data. We base our implementation on the visualisation proposed by \cite{Wongsuphasawat:2011}, focussing on answering questions regarding the number of records that include a specific event sequence, for example a question like  \textit{``has the patient already rejected a procedure or medication?''}  In \cite{EventFlow16}, the authors identified five main usability points in the medical domain which we have implemented accordingly: (1) visualise and review the data from individual records and their event sequences; (2) search for temporal patterns of interest, using a powerful graphical interface; (3) summarise all the event sequences, their timing and prevalence, and find anomalies; (4) perform data transformations to reveal useful patterns that answer questions you have; (5) select cohorts of interest for further studies.

In addition, \cite{Monroe:2013} provides a graphical approach to specify intervals and absences in temporal queries. Interval-based events represent a fundamental increase in complexity at every level of the application. In \cite{Malik:2016,Malik:2015}, the authors introduce a faceted search application by a special visualisation and manipulation interface. We adapted some key ideas e.g. concerning adjustments of events from this approach to our domain.

\section{Project Phases And Use Cases}
In this contribution we present a software system which was developed over the work on two medical use cases:
\begin{itemize}
	\item Use Case 1: \textbf{nephrology} patients, whose medical information is represented in the \tbase\ database of \charite\ Berlin, and
    \item Use Case 2: patients of UKE, the Universit\"{a}tsklinik Erlangen with medical data concerning examinations and their results with a focus on \textbf{mammography}. 	
\end{itemize}

The system was built from scratch for the first use case and later adapted to the second. The architectures of both systems are mostly the same, adaptations had to be made mainly in the indexing process and partly for the web interfaces for the clinician. Some functionalities are very dependent on the respective data situation and apply to one of the cases only.

\subsection{Use case 1: Nephrology}
The web-based electronic patient record \tbase\ was implemented in a German kidney transplantation programme as a cooperation between the Nephrology of \charite\ Universit\"{a}tsmedizin Berlin and the AI Lab of the Institute of Computer Sciences of the Humboldt University of Berlin \cite{schroeterTBase, lindemannTBase}. Currently, \tbase\ automatically integrates essential laboratory data (9.9 million values), clinical pharmacology (237.000 prescribed medications), diagnostic findings from radiology, pathology and virology (146.000 findings), and administrative data from the SAP-system of the \charite\ (70.000 diagnoses, 25.000 hospitalisations). 

Our very first text data set originated from the \tbase\ database containing medical information about nephrology patients. It consisted of about 5000 unstructured, free texts of four types:  ``Befunde'' (findings), ``Untersuchungen'' (visits), ``Entlassungsbriefe'' (clinical reports), and ``Verl\"{a}ufe'' (progress reports)\footnote{In this contribution we do not describe the first system version that handled only this part of the data as this can be regarded as an early proof-of-concept.}.
A much larger system with many additional features was built on further data from \charite\ about these patients: We worked on an extract of the original \tbase\ database containing 185 (anonymised) patients, including among others, their meta data, over 6300 diagnoses, 830,000 laboratory values, 25,000 medications, 12,000 examinations. The before mentioned medical texts of these patients were stored mostly in the data base, the clinical reports were stored externally as Word documents.

The main objective of this project phase was to realize a software system to support a physician in the search for patient cohorts. A focus was on the question how to present various types of patient attributes and medical facts in a coherent and intuitive way and how additional medical information hidden in unstructured data can be made accessible.

\subsection{Use case 2: Mammography}
In this subsequent use case the data was extracted from a proprietary radiology information system (RIS), a software system for managing medical imagery and associated data. We got large CSV files with examinations (``Untersuchungen'') findings (``Befunde'' and ``Beurteilungen''). Each of these files contains about 100,000 lines of patient records, each line built of both meta data about the patient and the examination and unstructured texts about the findings and evaluations. 
No further medical data about the patients was available.

Beside the possibility to have a sophisticated search over the data, the main focus was the question whether the realized solution of use case one would basically also fit to this differing data situation and how much effort it would take to adapt the modules.

\section{System Architecture}
\begin{figure*}
	\centering
	\includegraphics[width=1.0\textwidth]{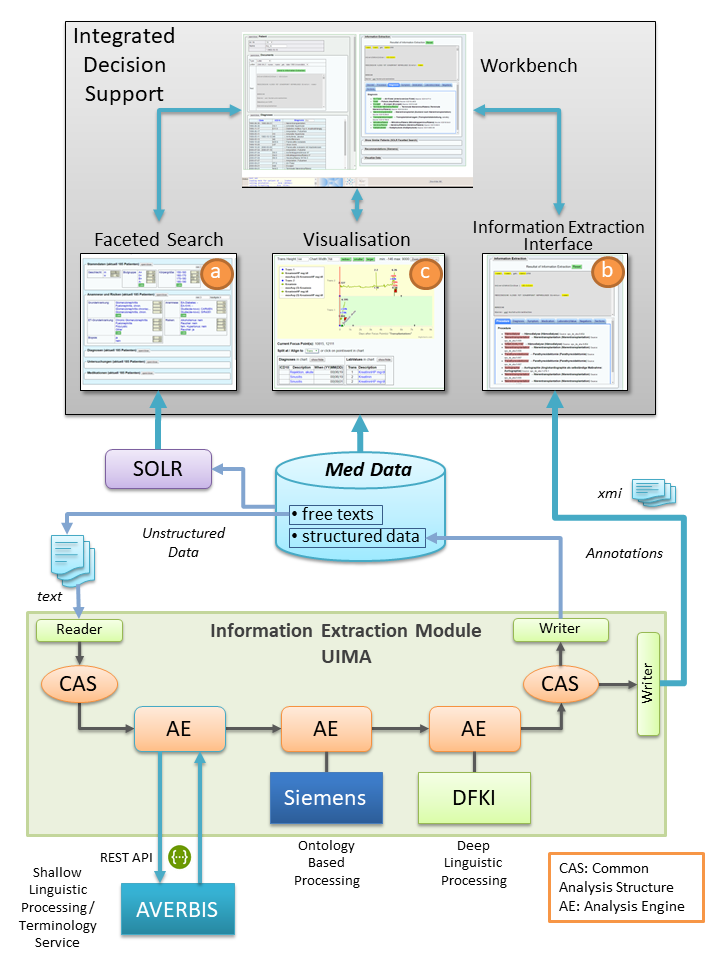}
	\caption{Basic System Architecture of the Integrated Information Extraction, Faceted Search and Visualisation}
	\label{fig:Architecture}
\end{figure*}

The main system (see figure \ref{fig:Architecture}) consists of two major components: (1) a backend with the medical data source and modules that support the search and information extraction on the medical data (lower part of figure \ref{fig:Architecture}) and (2) the web front end that enables the user to access all system functionalities in a single graphical user interface called ``Workbench''. The web interface combines three main modules: a) the faceted search on structured data as well as on automatically extracted data from documents, b) an interface to send text snippets (taken from the database or external sources) to the information extraction module and display results together with annotated text snippets, and c) an interactive visualisation tool to explore temporal relations between laboratory values and diagnoses. According to the specifics of the different use cases, variations of combinations of these modules were applied.

A typical medical database contains both structured data on patients in form of structured medical records (patient master data and facts about diagnoses, procedures, laboratory values, medication, etc.) as well as unstructured natural language texts in the form of clinical reports or findings.

The faceted search is built upon the Solr search platform\footnote{lucene.apache.org/solr/}, an open source enterprise search platform used in many large websites and applications. It is one of the most popular enterprise search engines.\footnote{db-engines.com/en/ranking/search+engine} Solr runs as a standalone full-text search server and uses the Lucene Java search library at its core for full-text indexing and faceted search.  
We chose the Solr system mainly because of some interesting features to faceted search, namely a proprietary query language that supports structured and textual search (cf. the aforementioned recommendations for faceted search interfaces \cite{HearstFacettedBrowsing} to include the support for a seamless integration with keyword search on the full text), its scalability and extensibility through plug-ins, and its various APIs for input (text, XML, JSON, Word, etc.) and output (JSON, XML, PHP, Python). A detailed description of our implementation can be found in section \ref{sec:IE}.

The information extraction is  performed by using UIMA  \cite{OASIS:UIMA:2009}, a framework of software systems that analyses large volumes of unstructured information in order to discover and annotate relevant knowledge. A UIMA pipeline is used to extract medical data about diagnoses, symptoms, medications, and laboratory values from the documents. The central data structure is a so-called CAS object (common analysis structure), which is guided through an analysis pipeline. The modules of this pipeline, called analysis engines (AE), enrich the CAS with additional annotations (lower part of the figure).

The results of the information extraction steps can be written back to the medical database as structured data, or they can be made available to the search module via the UIMA-Solr-API\footnote{wiki.apache.org/solr/SolrUIMA}\footnote{This very much depends on the respective data situation and access privileges.}. Moreover, the annotations can be evaluated using a separate presentation and validation interface. This architecture enables the user to search and access both the information newly extracted from unstructured text documents and the structured data of the medical database in a homogeneous way. The source of the facts is accessible as additional attribute provenance information. Section \ref{sec:IE} describes our implementation in detail.

The web user interface was built using AngularJS 1.3\footnote{www.angularjs.org}, a JavaScript-based open-source web application framework mainly maintained by Google to address challenges encountered in developing single-page applications. It aims to simplify the development of such applications by providing a framework for client-side model-view-controller (MVC) and model-view-viewmodel (MVVM) architectures. 

\section{System Modules And Interfaces}  

In the following secton, we describe the processing steps and system modules developed for the first use case we worked on, the \tbase\ database of \charite\ Berlin. 
Differing aspects of the second use case, data from UK Erlangen, and additional deployment details are mentioned especially when necessary but described in detail in section \ref{sec:casestudy}. We will also expound how the user interface enables the interaction with the system modules.

\subsection{Textual Information Extraction}\label{sec:IE}

The first step in our processing chain is information extraction.
One source of text data  originates from the \tbase\ database holding medical information about nephrology patients. It also contains about 5000 unstructured German texts. In a first preprocessing step, these texts are  processed by the project partner Averbis, who anonymises them and adds annotations to lexical items based on several medical reference systems and dictionaries (described in detail in section \ref{sec:casestudynephro}).

The text snippets that are sent to the information extraction module may be taken from the medical database or entered manually and in real-time, for example when a patient is referred from the family physician. Information from free text entries (i.e., text snippets that are typed in as opposed to pre-formulated labels chosen from menus) is especially interesting as it may contain negative facts (for example negative findings or rejected procedures or medications). In contrast to this, medical databases often contain only positive facts. A substantial fraction of the observations made by clinicians and entered into patient records are expressed by means of negation or by using terms which contain negative qualifiers (as in ``absence of pulse´´ or ´´surgical procedure not performed´´). This seems at first sight to present problems for ontologies, terminologies and data repositories that adhere to a realist view and thus reject any reference to putative non-existing entities.


The process of information extraction integrates existing external (Averbis) as well as internal local annotation modules to ensure state-of-the-art annotation results \cite{DBLP:journals/nle/KlueglTBFP16, DBLP:conf/cbms/BretschneiderZH17}.
Our pipeline starts with shallow linguistic processing combined with text mining modules to recognise diagnoses, medication, laboratory values, and other basic medical concepts  \cite{sonntag03, DBLP:conf/cbms/BretschneiderZH17}. 
Subsequently, additional analysis processes such as ontology-based semantic annotation and deep linguistic processing steps follow to recognise negations and their correct scope, i.e., dependency parsing \cite{DBLP:conf/clef/MkrtchyanS14}.
At the end of the pipeline we have an enriched CAS object that can be made available to the faceted search component via a custom interface (UIMA-Solr). Other CAS-consumer modules could be added that write the extracted facts  into the medical database as structured data or serialise them in different output formats (e.g., XMI or JSON).
This annotation module which serves the user interface is implemented as a Java servlet that takes text strings, sends them to the information extraction pipeline in the backend, and returns the annotations in the desired format.

\subsection{Presentation and Validation Interface}\label{sec:presentation_validation}

\begin{figure}
	\centering
	\includegraphics[width=1.0\columnwidth]{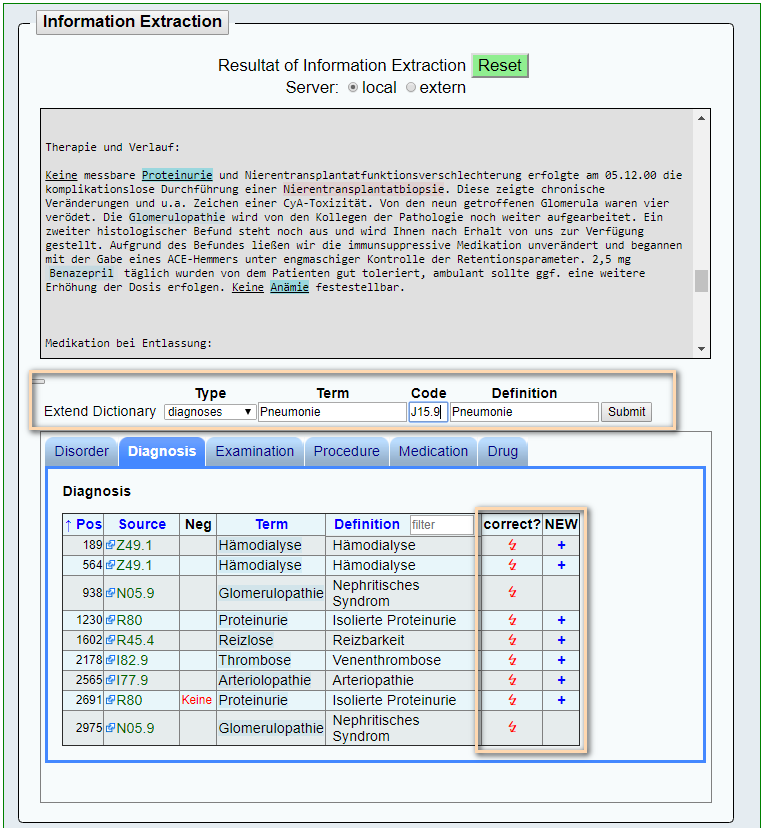}
	\caption{Presentation and validation interface with annotated text segments (upper part) and extracted results for disorders, diagnoses, examinations, procedures, medications, and drugs (lower part). 
	}
	\label{ieResults}
\end{figure}

The user frontend for an explicit annotation of a single document (in contrast to batch-processing large amounts of documents) includes  a presentation and validation interface (see figure \ref{ieResults}) that consists of two main parts: the upper part can be used to enter text and shows, after the annotation process has finished, the original text with highlighted annotations. The lower part contains tabs that list relevant annotations \footnote{They are obtained from the backend in XMI format (XML Metadata Interchange) and can be visualised in a pop-up by clicking on a highlighted word.}.  The goal is to create a convenient interface, which fulfils the clinician's needs.  Accordingly, this graphical user interface as Web page (this interface is part of the workbench shown in figure \ref*{fig:diffDbText}) serves three different purposes: (1) the display of unstructured source text documents, (2) the presentation of the original text snippets (source text part) found by the information extraction together with the extracted information, and (3) the validation of the annotations.

The presentation and validation interface has been developed after discussing with the clinicians the goals, behaviour patterns, and paper-based model scenarios of their interactions with results of an automatic information extraction process.  
Several guidelines for the relevant operating systems, which we want to adjust our interface to, are covered by the HTML5-based rendering engine of Solr. Firstly, the interface has to be functional. Secondly, it has to be convenient to the extent that it does not irritate the user in frequent work. For adaptability of the information extraction process in connection to validation procedures, we implemented the following functionality and frequent interaction patterns accordingly (they apply to different use cases according to the respective data situation):
\begin{itemize}
	\item If a relevant fact was not recognised because an item in one of the underlying dictionaries (diagnoses, disorders, examinations, procedures, or drugs) was missing, the user can add this item to the dictionary by using the input fields above the listed annotations (upper highlighting frame in figure \ref{ieResults}). The item will be added to a separate part of the dictionary and the UIMA pipeline will be restarted with the extended analysis engine. This allows both a direct extension of the extraction process by the user and the protection of the original dictionary\footnote{This was implemented in the second use case only because here we could control the whole annotation process including the underlying dictionaries.}.
	\item If an annotation is found to be incorrect, the clinician can click the red arrow in the second last column titled `correct?' (lower framed region). This triggers a message to the backend where these messages can be collected in order to check and adjust the annotation process via interactive machine learning\footnote{Here we implemented just the user interface to indicate an additional but not yet realised feature.} 
	\item If the text snippet is directly associated to a patient and the medical health record of the patient is available to the system (see also descriptions of figure \ref{fig:diffDbText} in section \ref{sec:casestudynephro}), this interface can also be used to add extracted facts to the record. The rightmost column titled `NEW' contains a link (the plus-sign) if the annotated term  is not yet contained in the health record of the database (see lower framed region in figure  \ref*{fig:diffDbText})\footnote{This is only applicable when additional medical data about a patient is available, i.e. in the first use case based on data from \tbase\.)}.
\end{itemize}

\subsection{Faceted Search}\label{sec:facetedsearch}

The faceted search uses Solr to (1) index available structured data and (2) enable the direct free text search on specific text fields. 
The patients' meta-data (structured medical data from databases or other sources) are modelled and indexed as main "patient objects" in Solr  with sub-objects for diagnoses, medications, examinations, etc. 
This object-oriented programming approach enables us to construct much more complex search queries\footnote{In fact queries with more than one restriction on a facet, e.g. \textit{`patients with laboratory test 1) for creatinin and 2) results over value x'} would not be possible without this hierarchical model. An example for a complex query can be found in section \ref{sec:timeeffort}.}. We can distinguish between predefined facets built  upon more simple attributes (e.g., gender, age, blood group, existence of findings) and complex dynamic facets (especially temporal relationships between events) which are assembled by the user while executing the faceted search. 

The faceted search works by providing information about existing values and the expected number of results  in real-time. This works even before an attribute (e.g., `examination of an organ') is restricted to a value (e.g., `thorax'). This feature is of great importance for one of the main applications, namely the identification of patient cohorts for medical studies. Additionally,  users may un-click any restriction made within this subsequent search process at any time, thereby giving them great flexibility in narrowing down and extending the search.

\begin{figure}
	\begin{mdframed}
		\centering
		\includegraphics[width=1.0\textwidth]{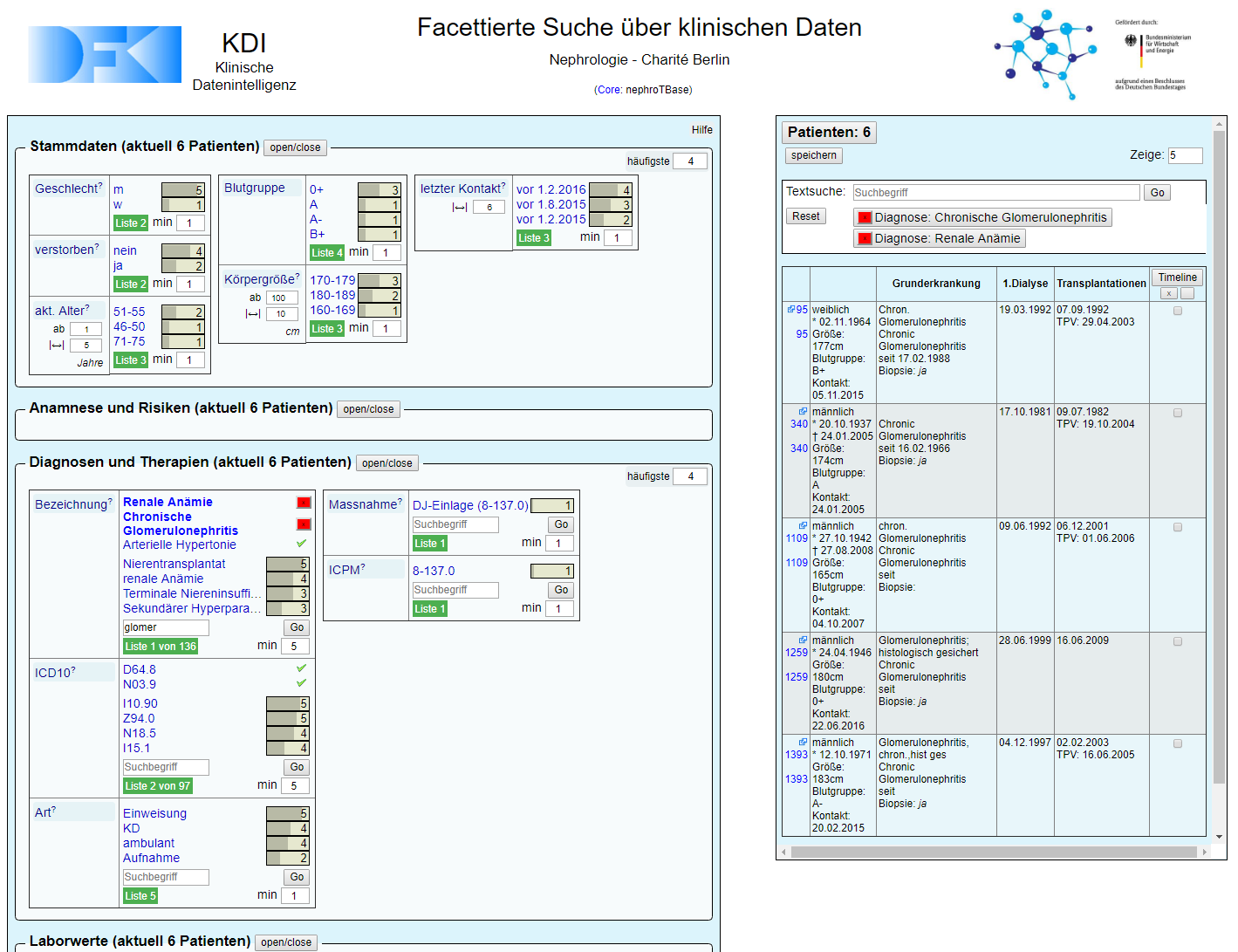}
	\end{mdframed}
	\caption{Web interface of Faceted Search with blocks of facets (left) and short presentation of 185 patients as result set (right).}
	\label{fig:webInterface}	
\end{figure}

\begin{figure}
	\centering
	\includegraphics[width=0.7\columnwidth]{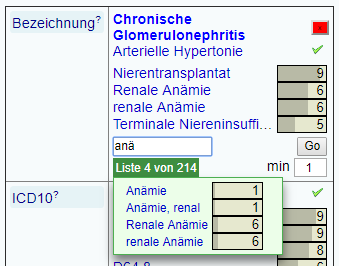}
	\caption{Presentation structure of textual facets}
	\label{fig:textualfacettes}
\end{figure}

The left part of the user interface in figure \ref{fig:webInterface} groups facets thematically in blocks like `Stammdaten' (patient master data), `Diagnosen und Therapien' (diagnoses and therapies), or `Anamnese und Risiken' (anamnesis and risks) that can be opened or closed\footnote{In this example, the facets in block `Stammdaten' are `Geschlecht' (gender), `verstorben' (deceased), `akt. Alter' (current age), `Blutgruppe' (blood group), `K\"{o}rpergr\"{o}\ss{}e' (body height), and `letzter Kontakt' (last contact).}. Each block  may contain several facets and each facet (see also figure \ref{fig:textualfacettes}) shows (a) current restrictions (bold face with delete-button), (b) the most frequent values with their cardinality, and (c) highlights the values that are  common to all remaining patients as well (marked with a green ``OK''-sign). This results in a maximum of information about a facet and the distribution of its values.

Additionally, there are lists with all remaining values (see green pop-up in figure \ref{fig:textualfacettes}) to allow the user to access all -- and especially -- less common values. This list  may be filtered directly by a completion mechanism: entering text in the field above the list (string `an\"{a}' in the example) restricts the list to entries containing this text snippet. This  is very helpful when dealing with different notations resulting from free text input for values (the list shows three variants for the diagnosis `renal anaemia': ``An\"{a}mie, renal'', ``Renale An\"{a}mnie", and ``renale An\"{a}mnie''). Additionally, using this input field the user can perform a free text search (including wild-cards) over the textual values of the facet (in the illustrated case in figure \ref{fig:textualfacettes}, the `term' attribute of diagnoses, called `Bezeichnung` in German).

Additional features include interval boundaries for numerical values that can be changed directly in the web interface (with immediate feedback, i.e., the presented intervals and their hit counts will change on the fly) or facets to indicate the last known correspondence with a patient (see for example the facets ``akt. Alter'' (current age) and ``letzter Kontakt'' (last contact) at the top of figure \ref{fig:webInterface}). These data are requested and retrieved from the backend (the Solr core), processed and transformed by the web client to build the lists, and displayed in menus of facet values as described above.
This extensive yet easily understandable organisation and presentation of facets and their values results in a maximum of control over the search process and a very flexible navigation within the facets (cf. the recommendations in \cite{HearstFacettedBrowsing} concerning facet navigation and exposure, graphic design, and keyword search).

\begin{figure}
	\centering
	\includegraphics[width=1.0\textwidth]{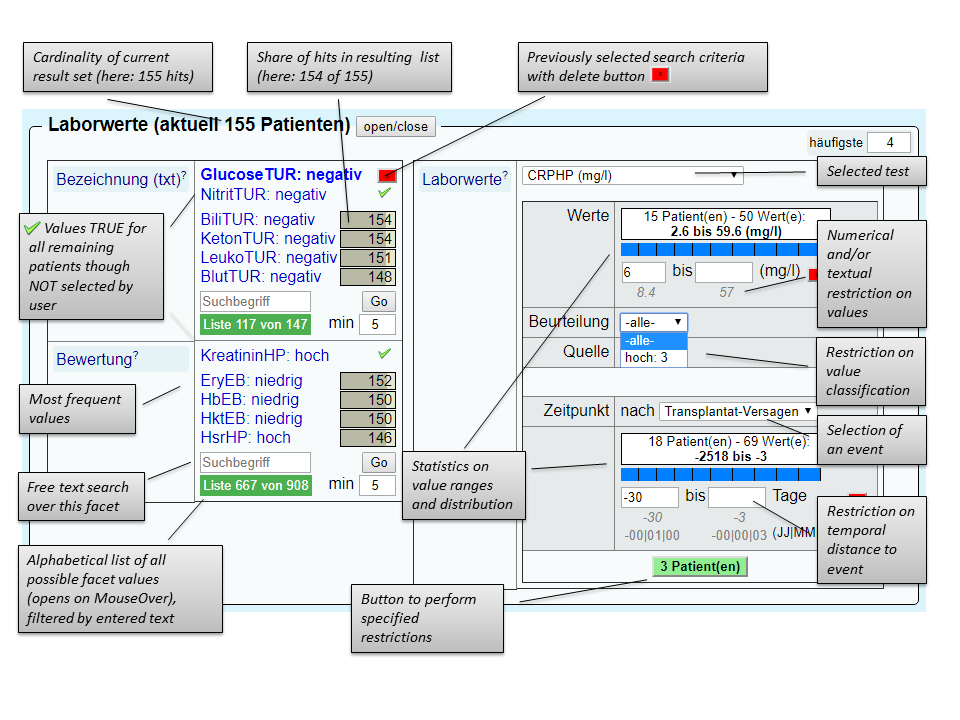}
	\caption{Definition of facets with textual values and classifications of laboratory results (left) and the specification of multiple restrictions on laboratory values (right)}
	\label{fig:laborwerte}
\end{figure}

The right part of figure \ref{fig:webInterface} consists of three elements (from top to bottom): (a) the possibility to save the result set (button `speichern' - save), (b) an input field to perform a direct search over some text fields (more details in section \ref{sec:freetextsearch}), and (c) a listing of matching patients as short profiles containing some meta data, basic diseases (`Grunderkrankung'), date of first dialysis (`1. Dialyse'), and some facts about transplantations and failures (`Transplanatationen').

\subsubsection{Dynamic Faceted Search Example}

In the following, we describe the specification of dynamic facets by the example of laboratory values. Figure \ref{fig:laborwerte} illustrates various aspects of this user interface. These facets consist of a term, the timestamp and the tested value, a textual or numerical result, an optional classification of the result and, as in our case, the provenance of the data (namely the structured database or the information extraction pipeline). Simple restrictions on these facts can be specified in the usual way in the left side of the corresponding block. 
But often these laboratory values are only interesting in combination with special events such as transplantations, failures, or rejections. To address this, the system allows for the temporal restriction of laboratory tests\footnote{As mentioned above, this kind of restrictions is only made possible by the complex hierarchical modeling of objects in Solr.} in relation to some predefined events and by that, to answer questions like \textit{``all patients with a value for `CRPHP (mg/l)' over 6mg/l, at most thirty days before a failure occurred''} (see right side of figure \ref{fig:laborwerte}).

In a similar way, the clinician  may also search for temporal restrictions between multiple endpoints such as basic disease, first dialysis, transplantations, rejections, failures, or death. For example, a query could ask for the result set of \textit{``all patients where the first rejection was within three days after the first transplantation.''}

As we have to deal with lots of facets and values and, more general, with great amounts of data, the user interface allows for opening and closing of facet blocks and by that for concentrating on facets in focus while hiding the rest. This also supports the performance of the web interface as only data of opened blocks is requested from the backend and included in the web page. With this functionality, the user is able to get an overview of the distribution of attribute values and may recognise correlations between chosen restrictions and other attribute values (e.g. most common medications within a given group of patients).

\subsubsection{Free Text Search}\label{sec:freetextsearch}

In addition to searching over structured data via facets, the user may want to perform a search for a specific attribute of patients that is neither covered by structured data nor annotations on texts. To address this request, the web interface allows to search for arbitrary terms in texts directly (see text input field in right upper part of figure \ref{fig:webInterface}). Which text is searched depends on the individual application---in \tbase\ external Microsoft Word documents are indexed, in UKE different parts of the examination data (``Befunde'', ``Beurteilungen''). Solr's text search also allows the user to use boolean operators and wild-cards which is very useful when dealing with free text searches entered by physicians and medical staff.

\subsection{Information Visualisation}

\begin{figure}
		\centering
		\includegraphics[width=1.0\textwidth]{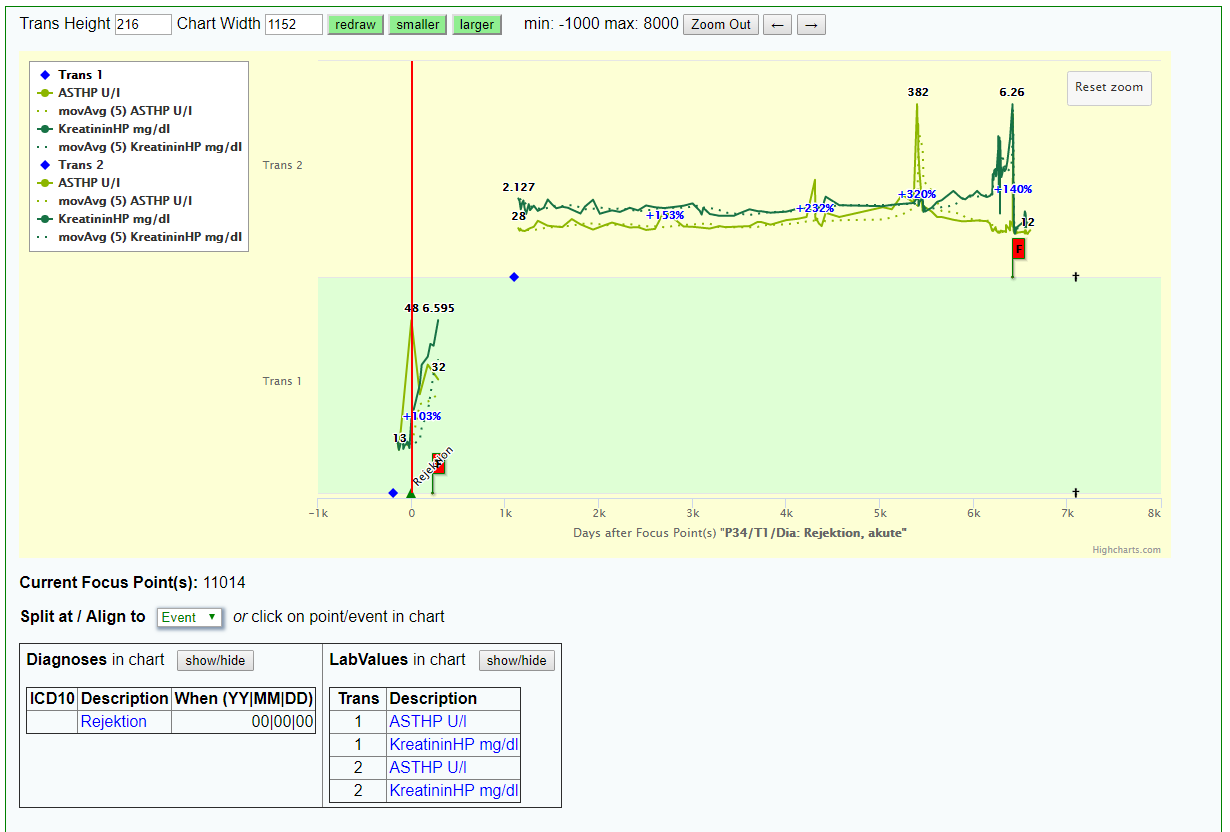}
	\caption{Visualization of temporal relationships between diagnoses and laboratory values (event sequences)}
	\label{fig:chart}
\end{figure}

\begin{figure}
		\centering
		\includegraphics[width=1.0\textwidth]{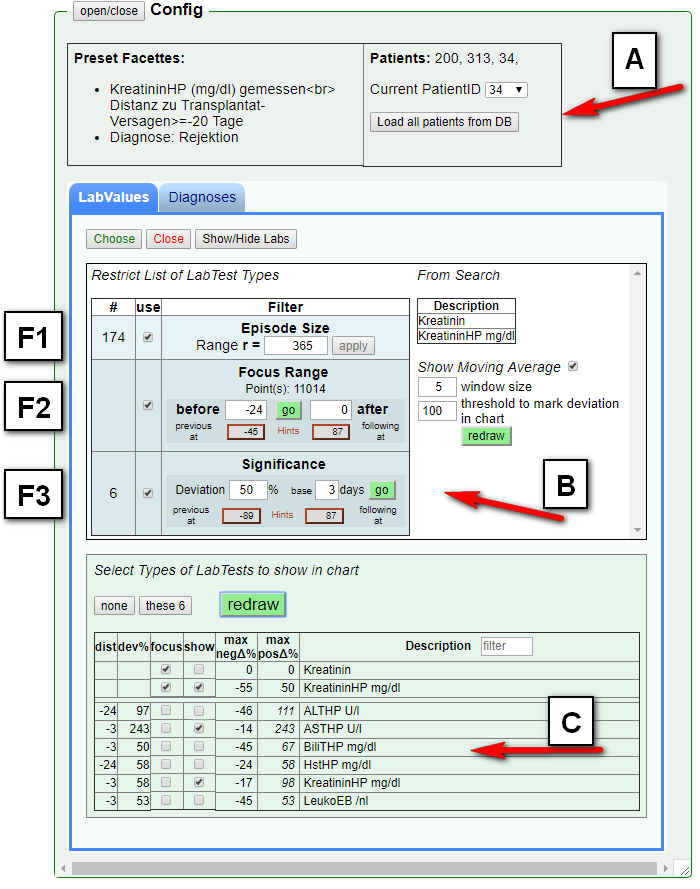}
	\caption{Configuration of chart contents with various filters}
	\label{fig:chartconfig}
\end{figure}

In contrast to other applications, in our visualisation application (broad structured data with timestamps from \tbase) the events (diagnoses and laboratory tests) are not predefined but can be determined directly from the attributes of the faceted search application, thereby providing a dynamic interactive tool. A special aspect of our domain is that each patient may have hundreds of laboratory tests with thousands of values. Therefore we focus on means to support the user in finding (potentially) relevant events and relations between them---for a single patient.
Figures \ref{fig:chart} (chart with some patient data) and \ref{fig:chartconfig} (corresponding configuration) display the two parts of the  visualisation interface.


The chart\footnote{drawn with highcharts (www.highcharts.com)} splits the data  by transplantations, i.e., each transplantation and its associated data are displayed in a separated horizontal layer of the chart, starting with the first at the bottom and the following stacked over it. It's essential that these stacked layers are always aligned to the date of an event, the current \textit{focus point(s)}\footnote{There can be multiple focus points at one time if the chart is aligned to the date of transplantations or failures.}: At the beginning, the alignment is done to the day(s) of the transplantation(s), but the user interface allows to change the alignment to any displayed event (including diagnoses and even each single laboratory value) just by clicking on it. The X-axis shows the distance of all events as number of days before or after the focus point (the actual time points given in date specifications would be of no use, especially as we want to stack the transplantations to be able to compare their chronological progressions). By marking an area with the mouse the user can zoom in and enlarge a time span to inspect it in more detail.
The chart part also shows some annotated values and (at the bottom) a list of the displayed data (hover with mouse highlights the corresponding items in the chart). The chart in figure \ref{fig:chart} is the result of the example configuration given at the end of this section.

The configuration part (figure \ref{fig:chartconfig}) contains information transferred from the search, namely selected patients and search restrictions over diagnoses and laboratory values in the upper region (area A). 
As the vast number of events (every single laboratory test value is a single event) could not be displayed reasonably in a chart, we implemented different kinds of filters over \textit{types} of diagnoses and laboratory tests that support the user in restricting  (1)  the events to a manageable amount and/or (b) to some interesting and relevant period of time. Area B in figure \ref*{fig:chartconfig} shows the interaction options with the different filters, their adjustment elements and the resulting amount of hits (in the first column). The lower part (area C) contains a list of those types that have passed all applied filters with some additional information, and options to include them in the chart or to search for a specific description term (\textit{Term Filter}).

\subsection{Filters for Diagnoses and Laboratory Tests}

\begin{figure}
	\centering
	\includegraphics[width=0.6\columnwidth]{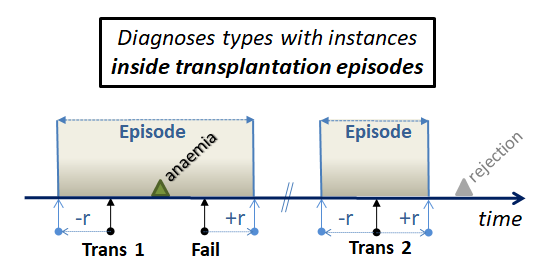}
	\caption{``Transplantation Episode Filter'' (Filter F1 in \ref{fig:chartconfig}): transplantation episodes with range \textit{r}}
	\label{fig:filtersEpisodes}
\end{figure}

\begin{figure}
	\centering
	\includegraphics[width=0.6\columnwidth]{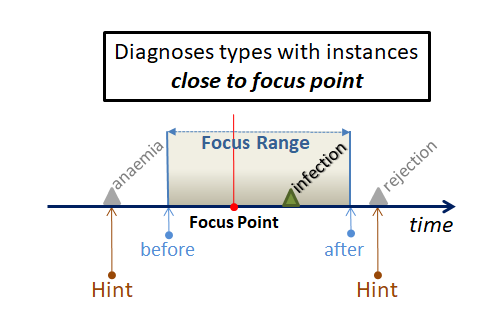}
	\caption{``Focus Range Filter'' (Filter F2): ranges around focus points}
	\label{fig:filterFocusRange}
\end{figure}

\begin{figure}
	\centering
	\includegraphics[width=0.6\columnwidth]{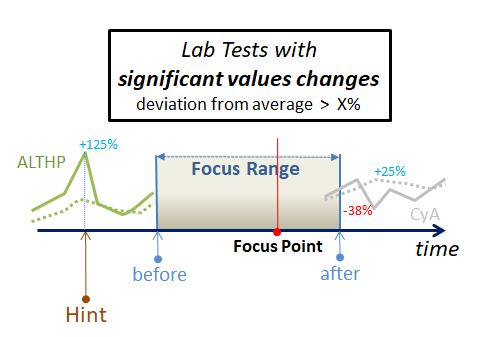}
	\caption{``Significance Filter'' (Filter F3): visual clues to significant value changes}
	\label{fig:filterSignificance}
\end{figure}

In this section we describe the different filters over diagnoses and laboratory tests. In Figure \ref{fig:chartconfig} area B points to the filter configurations, figures \ref*{fig:filtersEpisodes} to \ref*{fig:filterSignificance} illustrate the individual filters:
\begin{itemize}
\item F1: ``Transplantation Episode Filter'' (figure \ref{fig:filtersEpisodes});
\item F2: ``Focus Range Filter'' (figure \ref{fig:filterFocusRange});
\item F3: ``Significance Filter'' (figure \ref{fig:filterSignificance}).
\end{itemize}

The first filter is independent from the other two, and each filter can be activated individually by the check boxes (column ``use'' of the configuration). 
	
Filter F1 is derived from the fact that not all events in the potentially long patient record (which may span entries over decades) are of interest. Data of relevance is normally located near (or between) some specific events. In the nephrology domain, these events are transplantations and failures. We associate the data to so-called \textit{transplantation episodes} which start at some given time span  before the transplantation date and last until this span after a failure or the death of the patient (as figure \ref{fig:filtersEpisodes} illustrates), or the transplantation itself, if neither of them happened. The time span (named range $r$) of filter F1 can be adjusted in the web interface as ``episode size''. This filter narrows all data and events down  to those \textit{``near the transplantation of a kidney and the potential loss of the same''.}

Filters F2 and F3 take into account that the chart is always aligned to some event, the focus point. It is likely that the user is interested in other events that happened in a time region around this time point which we will call the ``focus range''.

Filter F2, the \textit{Focus Range Filter}, allows us to restrict events by their distance to the focus point (see figure \ref{fig:filterFocusRange}). As the user does not know in which time window or distance the next events are located, the system takes over the search for them (within the available patient data) and suggests both a distance before and after the focus point, realised as dynamic buttons (red buttons in figure \ref{fig:chartconfig} that fill their values (labels) into the input fields above them when clicked). These nearest distance values are dynamically recomputed each time the range changes allowing the user to enlarge the range step by step, starting from the focus point itself.
Every time the focus range is changed, the list of event types shown at the bottom is automatically updated.

Even with the two filters described above, it is sometimes hard to find {\it significant} events such as a sharp increase in a laboratory value. Drawing multiple laboratory lines in a chart at the same time makes it rather confusing very quickly. It would be more helpful if the system searched for major changes in the course of a line and presented  clues about the nearest of these changes. We integrated such a search for major changes as a third filter F3 which only applies for laboratory tests, not diagnoses (see figure \ref{fig:filterSignificance}). F3, the \textit{Significance Filter}, computes baselines (or the `moving average') of laboratory values (the average value over a certain period of time) and compares each single value to it. The user can specify the size of the baseline ``window'' (days) and the percentage of change he or she is interested in. Similar to F2, buttons show the hints and their labels are recomputed after every change. The baseline can be integrated in the chart as a dotted line coupled with the laboratory line (configured in the right part of area B in figure \ref{fig:chart}).

A possible use of these tools and functionalities is shown in the following scenario (figure \ref{fig:chart} shows the resulting chart): consider the user is interested in whether or not there are significant changes in some laboratory values of a patient shortly before a rejection occurred. To find out about this, she can do the following:

\begin{itemize}
	\item select diagnosis type ``Rejection'' (by selecting the ``Diagnoses'' Tab, using only the episode filter and the term filter);
	\item redraw the chart and align it to a rejection event (this event is now the focus point);
\end{itemize} 

\begin{figure}
	\centering
	\includegraphics[width=1.0\columnwidth]{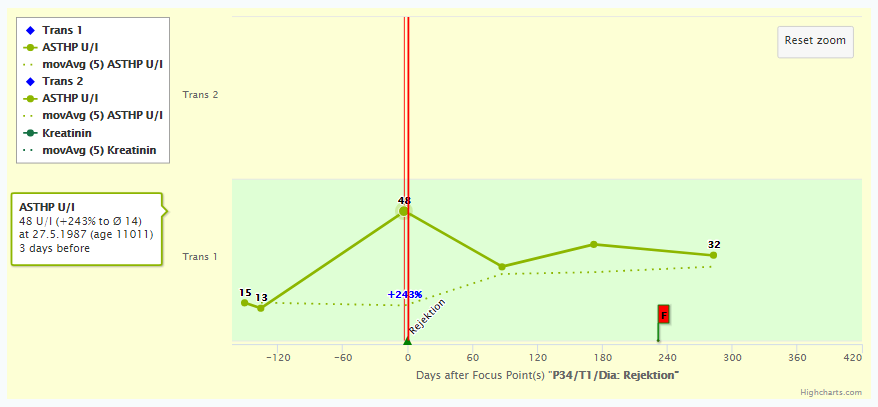}
	\caption{Chart showing a significant value change in some laboratory lines just three days before a rejection was diagnosed}
	\label{fig:significant}
\end{figure}

In the ``LabValues'' tag,  F1 now shows a list of all laboratory tests that have values for the date of the focus event. These may be a list of hundreds of tests. The hint-buttons of F3 show where a laboratory value can be found that satisfies the given significance thresholds. Then she may 
\begin{itemize}
	\item click on the left red button to expand the focus range and get a list of laboratory test types together with their highest deviation values and when they were taken;
	\item choose some types of tests and trigger the redrawing of the chart.
\end{itemize} 
The chart now shows a picture similar to that in figure \ref{fig:significant}, where the tested value of ``ASTHP U/I'' showed a significant rise of 243\% to 48U/I just three days before the rejection was diagnosed (a tool-tip is shown with a green border on the left side horizontally levelled with the value).

\section{Case Studies and Deployment Details}\label{sec:casestudy}

Two use case studies in nephrology and mammography are presented to illustrate the usefulness of the faceted search and visualisation as an integrated decision support application. In these case studies we explore how our architecture of open-source tools combining textual information extraction, faceted search, and information visualisation supports the task of exploratory data analysis and the understanding of faceted search results for cohort selection. Through interviews with domain experts, we iteratively designed and implemented both the interactive faceted search and visualisation. We investigate the strengths and limitations of our application as a cohort selection tool. We worked with medical researchers.  In the informal clinical evaluation, we observed two senior experts of nephrology of the clinical department and radiologists with experience in breast imaging in the mammography scenario with real patients, respectively. Additionally 6 medical experts tested the faceted search applications apart from the daily routine. These experts also controlled the evaluation of the task-based  questions. 
After several interviews to understand the clinician's goals and questions for both use cases, we observed them as they conducted the search on traditional SQL driven database tools. Specifically, the medical researchers' questions were: 

\begin{itemize}
	\item Questions concerning data quality, correctness and completeness
	{\begin{itemize}
			\item Are there correlations between attribute values? (to be answered by faceted search and/or visualisation)
			\item Are there similarities in specific aspects of medical records? (visualisation)
			\item Are there contradictions in structural data and texts (information extraction, followed by a comparison with structured data)
			\item Is there additional information in unstructured texts (e.g., from transferring physician)
	\end{itemize}}
	
	\item Questions concerning planned medical measures like examinations, medication
	{\begin{itemize}
			\item Has the patient already rejected some procedures or medications? (only documented in textual descriptions)
			\item Have previous interventions been successful?  (only documented in textual descriptions)
			\item Which similar patients have been treated  with success (faceted search)
	\end{itemize}}
\end{itemize}

The case studies are based on qualitative assessments of our system against recent guidelines for systematic exploration of event sequence (see figure \ref{fig:chart}) comparisons \cite{Malik:2016,Malik:2015}, i.e., (a) reduce wait times during computation; (b) convey hypotheses succinctly; (c) visualise statistical results and differences; (d) allow flexible methods for organising results; (e) provide flexible interactions for parsing results. These guidelines were provided to the clinicians and function as a basis to a systematic evaluation.
Overall, the initial feedback was positive and the clinicians of both use cases valued the possibility to narrow down the search space by clicking on easily-to-comprehend dynamic facets that we provide. We also determined whether the results were consistent with their expectations. First results suggest a reduction of search time by 80-90\% on structured values for cohort selection. First results of displaying results for ``has the patient already rejected some procedures or medications'' suggest a 95\% reduction in the time needed to find the corresponding textual description in the documents.  
More details are provided in the next two subsections. 

\subsection{Use Case 1: Nephrology at \charite}\label{sec:casestudynephro}

Our faceted search application for this use case is based on the nephrology EHR database \tbase. The web-based electronic patient record \tbase\ was implemented in a German kidney transplantation programme as a cooperation between the Nephrology of \charite\ Universit\"{a}tsmedizin Berlin and the AI Lab of the Institute of Computer Sciences of the Humboldt University of Berlin. 

Our first text data set originates from the \tbase\ database of \charite\ containing medical information about nephrology patients. It consists of about 5000 unstructured, free texts (findings, visits, clinical reports, and progress reports. The free texts have been preprocessed by the project partner Averbis, who anonymises the texts and adds annotations based on several medical reference systems and dictionaries such as 	LOINC\footnote{loinc.org}, ICD10\footnote{www.icd-code.de}, ABDAMED\footnote{abdata.de/datenangebot/abdamed/}). The output of the information extraction process was serialized in XMI format and indexed with the so called \textit{XPathEntityProcessor}\footnote{lucene.apache.org/solr/guide/6\_6/uploading-structured-data-store-data-with-the-data-import-handler.html\#the-xpathentityprocessor} of Solr. As these texts did not contain any metadata about the patients, searches could only be made over the annotations.

A much larger system with many additional features was built on further data from \charite: We worked on an extract of the original \tbase\ database containing data about 185 (anonymised) patients.
Most of the examples and figures above used to illustrate the user interface and system functionalities are taken from this system version. The number of feature blocks thereby increased heavily (e.g., interval ranges for numerical values, dynamic facets like complex searches over laboratory values and endpoints, and a great number of utility functions to fine-tune the data processing for correct representation of data and information). 

In addition, the performance of the user interface moved into focus, as the amount of data to load, store, and render increased heavily. One of the key issues in this respect arises from the fact that almost all input from medical staff is made via text fields, which results in various different textual  formulations of the same medical concept.

\begin{figure}
\centering
	\includegraphics[width=0.65\textwidth]{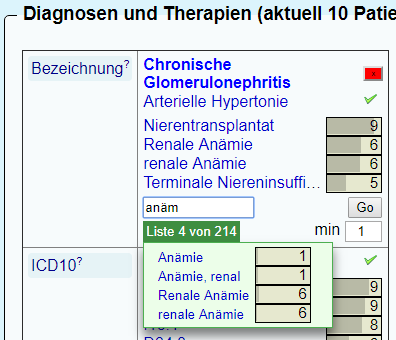}
	\caption{Restriction by using substrings of diagnosis terms: the menu is restricted to items containing the entered substring ``an\"{a}m''. Clicking on button ``Go'' will include all shown variants.}
	\label{fig:searchBySubstring}
\end{figure}

Two requirements had to be taken into account: (1) the web page must respond to user interactions quickly enough, and (2) the user should be given the best possible overview over existing textual formulations and variations to form exact filters. We came up with the following solution which satisfies both requirements:  Besides the four (this is the default) most frequent values of the hierarchical items from the current selection, the presentation of a textual facet should include a menu with an alphabetical list of these values (see figure \ref{fig:searchBySubstring}). It turned out that building this menus takes most of the time for the web page to respond after opening a block. The menu list can be restricted in two ways: (1) a minimal number of hits for a value or (2) by a substring that can be entered just above the menu. By means of this input field the user can inspect the variations of a facet by entering different search strings (e.g., the string ``an\"{a}m'' in figure \ref{fig:searchBySubstring} which restricts the alphabetical list to 4 out of 214  terms). As a result of this sort of filtering (instead of clicking on a link) the user triggers a search over all diagnoses that are matched by this string. 

Figure \ref{fig:searchBySubstring} also shows how the system visualises correlations between facets (or facet values): (1) the term ``Chronische Glomerulonephritis'' was previously clicked by the user (provided with a red button for deletion), (2) the term ``Arterielle Hypertonie'' was marked by the system with a green OK-sign to indicate that all remaining patients also received this diagnosis and (3) the list of the four most common remaining values together with their cardinality indicate decreasing correlations to further diagnoses.

A further issue of the \tbase\ data source was the fact, that the text documents that should be indexed are not stored in the research database itself. Instead, they are stored in the file system as Word documents and the database only contains a link. In order to include the texts in Solr, we made use of a special indexing feature, the \textit{TikaEntityProcessor}\footnote{lucene.apache.org/solr/guide/6\_6/uploading-data-with-solr-cell-using-apache-tika.html}.
This framework allows us to integrate external document types, including Word, PDF, and RTF. Accordingly, the text documents have been loaded into a separate Solr core ``Letters''. When indexed patients from \tbase\ are reqeusted, the text contents from the letters are retrieved from this letters core via a \textit{SolrEntityProcessor} which we implemented.

Typical requests and questions that can be posed and answered by the resulting faceted search include:
\begin{itemize}
	\item 
	Make a list of female patients between 50 and 60 years, with diagnosis ``Renale An\"{a}mie''.	
	Which other diagnoses occur most often? Save the list of patients.
	\item Make a list of patients with (1) a second transplantation at most 5 years after the first transplantation, (2) age at second transplantation at most 30 years. Which medication was administered most commonly?
	\item Make a list of short-term and long-term surviving patients with a laboratory value for ``KreatininHP'' exceeding 5mg/dl max 10 days before a transplantation failure. Are there any commonalities in other lab values for these patients?
	\item Make a list of patients with a suspicion on inflammation within the first three days after a transplantation. When did a failure occur with these patients? 
	Show the medical timelines.
	

\end{itemize}

\begin{figure}
	\centering
	\includegraphics[width=1.0\textwidth]{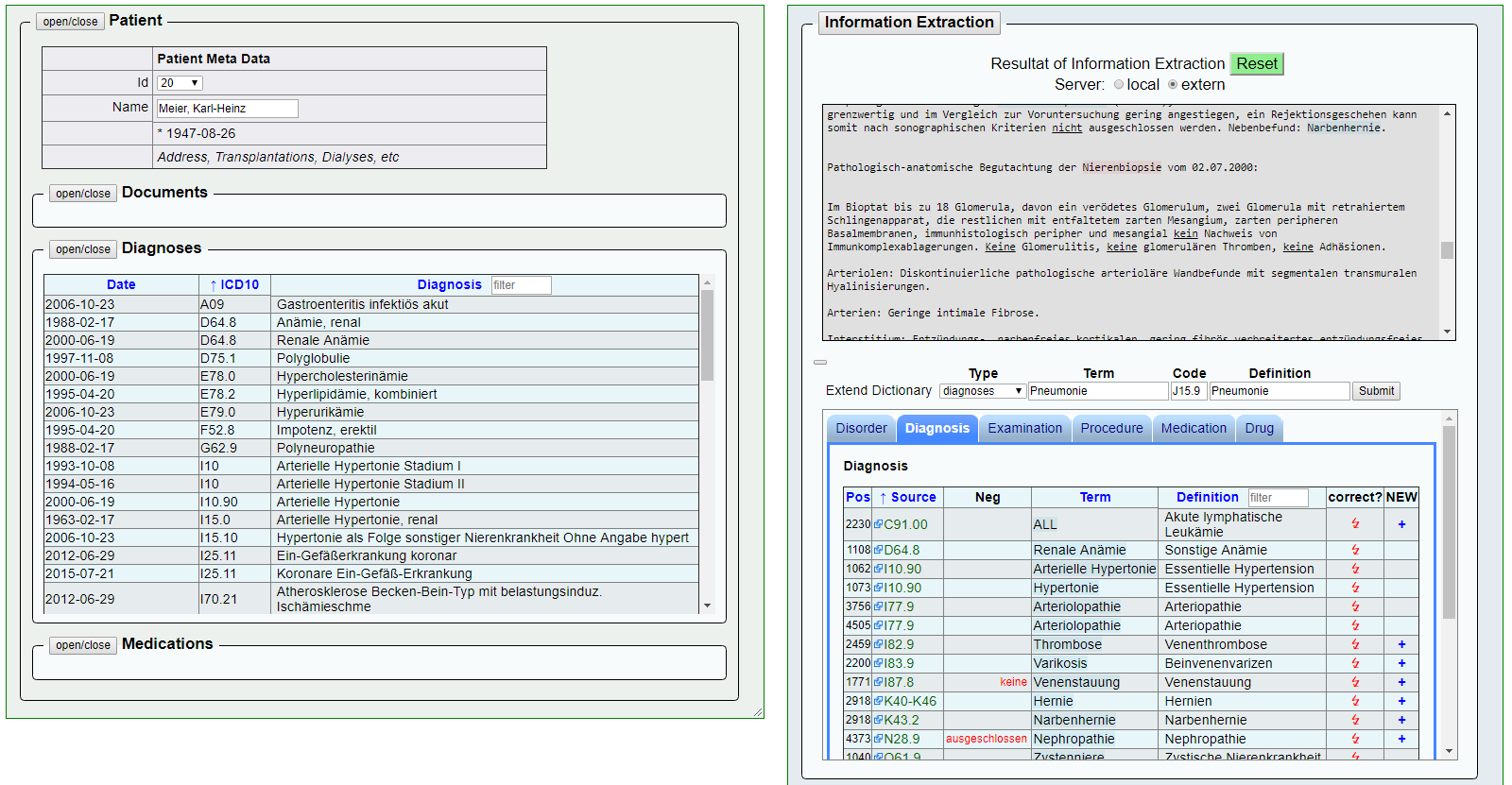}
	\caption{The physician workbench}
	\label{fig:diffDbText}
\end{figure}

Figure \ref*{fig:diffDbText} shows a screenshot of the whole workbench. The workbench is a web page integrating a direct access to clinical database contents like texts, diagnoses and medication (left side) plus the interface for the information extraction (IE) process  (right side) and links to other modules like faceted search and timeline (not visible in this truncated picture). 
By this a direct comparison of structured data and data contained in unstructured texts is made possible (e.g., previously unknown diagnoses are marked with a ``+'' in the last column named ``NEW'') and demonstrates the immediate benefits the user gets by using the workbench based on the IE pipeline:
\begin{itemize}
	\item IE may detect negated facts, which are not stored in the database\footnote{The results of the IE shows the single negating terms in a separate column.};
	\item unstructured texts (especially clinical reports) often contain additional information about the medical treatment not covered by the structured data (medical reference systems and dictionaries), for example procedures and medications proposed by a clinician but rejected by the patient. This information is made explicit in the annotated text from the IE process;
	\item the direct comparison of structured data in the electronic record and data found in textual documents may show confirming overlapping information,  contradictions between them as well as provide complementary information.
\end{itemize}

\subsubsection{Performance tests}

We made several performance tests with the Chrome browser, version 62.0.3202 for  Linux, macOS and Windows and present average performance times:
\begin{itemize}
	\item loading of start page, no open blocks: 550KB: 1500ms ; the block ``Diagnoses'' contains four main facets: diagnosis terms (2233 items), ICD10-codes (979 items), therapy terms (74 items), and therapy codes (51 items).
	\item opening Block ``Diagnosis'': response from Solr: 200KB after 250-300ms (TTFB),
	\item scripting and showing all menu items: 3000ms 
	\item showing only items with hit count $>$ 5 (2 menus with approx. 160 items): 800ms
\end{itemize}

In order to implement these real-time functionalities and to decrease response times after facet changes (i.e., to minimise the number of backend queries), we always request the complete list of values for text facets when a block is opened. The items that are actually shown in the menu are however restricted by the `mincount' (default 5) at the start and additionally by the substring the user has entered. This filtering can only be done on the web client without requesting the backend (thereby implementing the client-side model-view-controller (MVC)).
It should be emphasised that the amount of \textit{different} diagnosis terms is not proportional to the total number of patients (heavy tail distribution on new terms), and only these different terms are of concern for the search (exponential growth), not the number of patients which is linear in Solr's indexing mechanism. In the version installed at \charite\ with over 4000 patients and corresponding numbers of terms (around 3500), we set the default value for the minimum hits to 10 to achieve a rel-time response time of 1500ms.

\subsection{Use Case 2: Mammography at UK Erlangen}

In the use case ``UK Erlangen'' we had a very different data situation: the data was extracted from a proprietary radiology information system (RIS). We got large CSV files with the examinations (``Untersuchungen'') and findings (``Befunde'' and ``Beurteilungen'') on a timely basis. These files each contain about 100,000 lines of patient records, each line built of both metadata about the patient and the examination and unstructured texts about the findings and evaluations. 
Two main adaptations had to be made to cope with the different source, structure, and content of the data. In our first use case (\charite\ Berlin), the most interesting texts (in terms of extractable medical information) had been so called ``Entlassungsbriefe'', letters from the hospital to the family physicians describing the anamnesis, laboratory tests at the beginning and the end of the hospital stay and the course of treatment. The information process was tuned to give best results for this kind of texts. For the current use case we had to  add new information extraction analysis engines to cover different information in the texts, for example the recognition of BIRADS terms. 

\begin{figure}
	\centering
	\includegraphics[width=0.9\columnwidth]{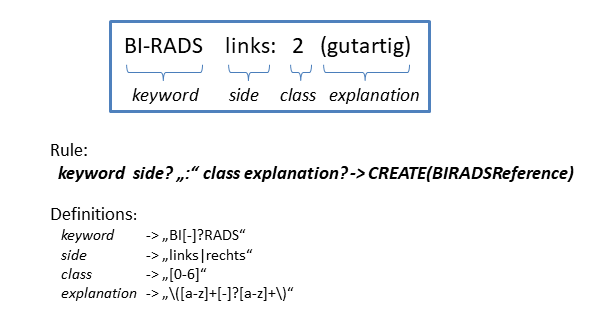}
	\caption{Simple RUTA rule (regular expression) to detect references to BIRADS classifications in findings.}
	\label{fig:rutaRule}
\end{figure}

This has been accomplished by integrating RUTA  (`Rule-based Text Annotation' \cite{DBLP:journals/nle/KlueglTBFP16}) as an additional text analysis engine and defining appropriate rules. The integration of the RUTA engine into the existing Java project of the information extraction pipeline turned out to be a complex integration step. We used the UIMA Ruta Maven Plugin described in the RUTA Book \cite{UIMABook}  and recommend the example project \footnote{svn.apache.org/repos/asf/uima/ruta/trunk/example-projects/ruta-maven-example} mentioned there. Figure \ref{fig:rutaRule} shows a simple rule to detect different forms of BIRADS classifications. 
Additionally, we added new annotation types like examination methods (e.g., sonography, mammorgaphy, MRT, or CT). These methods are explained in detail in \cite{DBLP:conf/cbms/BretschneiderZH17}.

Another adjustment was made necessary by Solr's indexing mechanism. 
The modelling of a patient and all his properties results in a hierarchy of depth three: level one, a patient has some metadata (e.g., name, address, date of birth), and level two, examinations which also have some metadata (e.g., date, physician, kind of examination, finding text and evaluation). The third and lowest level comprises of annotations within the texts (e.g., diagnoses, procedures, disorders, medications).
A Solr core can only handle one single database object, which in our case are patients. Sub-objects are only supported in a rather restricted manner. The problem was that an object must always be indexed together with all sub-objects. This is incompatible to the fact that (as described above) a patient may be examined more than once at different dates, and the documentation of the examinations is scattered over the source files (the lines of the files are ordered by examination times, not patients). 

If we updated the data of a patient with a second finding, we would overwrite the data of the first finding. We came up with a technical solution similar to the ``letter''-case in the \tbase\ domain; this solution can also be used in replicated architectures at other universities, research institutes, or hospitals: we use a second Solr core to only store the facts of findings, i.e., in this core the findings are the main objects. 
Further implementation details should be mentioned: the UIMA annotation pipeline generates two kinds of output for every finding: an XMI file, containing the data of a finding together with all annotations, and a text file containing the metadata of the patient. Solr first indexes the findings, storing them in the first core. After that, the main core is indexed, reading the patient files and requesting the first core to get all findings of a patient. The user interface only interacts with the main core whose main objects are patients. This procedure also allows later updates when new findings are present: after adding them to the secondary findings core the patient is re-indexed with all available findings.

\begin{figure}
	\centering
	\includegraphics[width=1.0\textwidth]{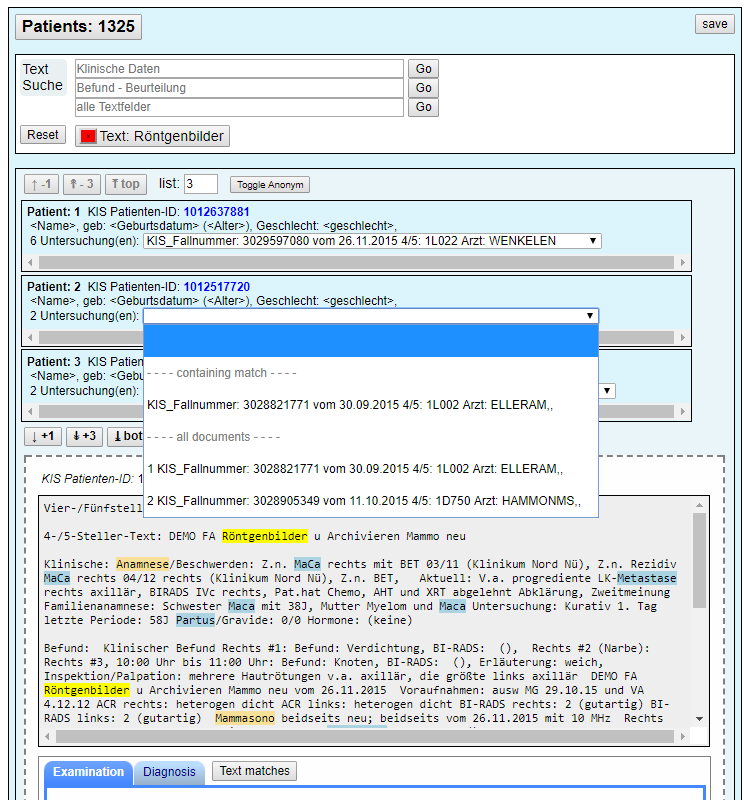}
	\caption{Presentation of patients together with texts of examinations}
	\label{fig:patientstexts}
\end{figure}

A second challenge in this use case became apparent when the search results were inspected by the  team of radiologists: the user searches mainly over the facts of findings, but the resulting documents are patients. If we presented only the metadata of patients as results, the user had no options to check any details of the results of his search. Therefore the part of the user interface that presents information about a hit was extended as can be seen in figure \ref{fig:patientstexts}: Each patient is presented with some metadata (here anonymised) and a menu with all texts of his or her examinations. When the user uses a free text search (in this case looking for ``R\"{o}ntgenbilder'', x-ray images), the menu marks the texts where the string was found (``\textit{-- containing match --}'') and highlights the corresponding passages in selected text (marked in yellow).

The information extraction engines built on top of the UIMA pipeline \cite{DBLP:conf/cbms/BretschneiderZH17} usually work on dictionaries of diagnoses, symptoms, medications etc. The different wordings used in the texts cannot be completely covered by general dictionaries, which means that the facts contained therein are not always recognised. In addition, clinicians and other health care professionals often use specialist jargon, and it would be desirable to enable a user to expand the underlying dictionary. 
Therefore we included an appropriate end-user functionality\footnote{This was implemented by us only in this use case because in the nephrology use case this part of the annotation process was realized by an external project partner.}: the presentation of annotated texts and extracted annotation types contains an input area, where the user can add an entry by choosing the type of dictionary and providing a term, (optionally) a code, and a definition. This data is sent to the servlet that  interfaces with the information extraction of UIMA, which adds the item to a separate part of this dictionary, recompiles the appropriate analysis engine and restarts the pipeline. The next time the information extraction is triggered, the new entry will be considered. 
By keeping new entries given by the user separate from the system dictionaries we facilitate an editorial revision mechanism for user entries.

\subsubsection{Performance tests}

We made some performance tests  (Chrome browser), version 62.0.3202 for  Linux, macOS, and Windows and present average performance times:

\begin{itemize}
	\item loading of start page, no open blocks: 1.1MB, 1000ms; the block ``Diagnoses'' then contains three main facets: diagnosis definition (1507 items), terms (1804 items), ICD10-codes (1405 items).
	\item opening Block ``Diagnosis'': response from Solr: 380KB after 250-300ms (TTFB)
	\item scripting and showing all menu items: 3800-4200ms 
	\item showing only items with hit count $>$ 10 (3 menus with approx. 550-650 items): 1.8-2.2s
\end{itemize}

\subsection{Time and Effort}\label{sec:timeeffort}

Table \ref{tab:efforts} shows an estimation of personnel costs for a senior software engineer to build the main parts of the presented software, with no experience in the deployed tools as Solr, UIMA, Angular, HighCharts, and with no local experts available. This is a crucial point as the documentation especially for Solr mostly only covers simple examples---details about building queries concerning sub-objects are very sparse. To get an idea about the complexity of the resulting queries, we give an example for the Solr request to get the values for diagnoses terms (shortened, the actual block contains four more facets)  after a temporal restriction of a laboratory value:

{\small
\begin{verbatim}
http://localhost:8080/solr/nephroTBase/select?wt=json
    &q=*:*
    &start=0
    &rows=5
    &facet.mincount=1
    &fq={!tag=DT}doctype:patient
    &fq={!parent which='doctype:patient' tag=FQ}
    _query_: {!parent which=doctype:labor}
              lab_description\_canon: kreatininhp_mgdl AND 
              labnumval: [* TO 10] AND
    _query_: {!frange u=500} sub(lab_days, epi_start_days)
    &json.facet={ dia_description: { 
                  type: terms, 
                  field: dia_description, 
                  limit: -1, 
                  domain:{ blockChildren: doctype:patient},
                  facet: { patients: unique(patient_id) }
                  }
                }
\end{verbatim}
}
\noindent 
\begin{table}
\begin{tabular}{|l|r|r|}
	 \hline 
	 & \multicolumn{2}{|c|}{personnel costs (days)} \\ 	\hline
													    	  & nephrology & mammography\\ \hline
	  
	\textbf{Solr}  &  &  \\ 	\hline
	 familiarisation with (aspects of) the topic					& 15 &  2 \\ 	\hline 
	 configuration of indexing procedures  							& 15 &  8 \\    \hline 
	 specification of schemata  									&  8 &  2 \\ 	\hline 
	 implementation of Solr requests/queries 						& 15 &  1 \\ 	\hline 
	 realization of complex facets				 					& 15 &  - \\ 	\hline 
	 inclusion of external sources (word documents)  				&  3 &  - \\ 	\hline 
	
	\textbf{UIMA} & &  \\ 	\hline
	 familiarisation with (aspects of) the topic					& 10 &  5 \\ 	\hline 
	 implementation of environment (servlet, interfaces)			&  5 &  1 \\ 	\hline 	 
	 dictionaries, pre-compilation 		 							&  - &  8 \\ 	\hline 	 
	 AEs for readers, word splitting, stemming, etc.  	 			&  - & 10 \\ 	\hline  	 
	 integration of RUTA Engine, definition of rules  	 			&  - & 10 \\ 	\hline 	 
	 visualisation of annotation data 				   	 			& 10 &  2 \\ 	\hline 	 	 
	
	\textbf{Angular} & & \\ 	\hline
	 familiarisation with the topic 								& 10 &  - \\ 	\hline 
	 implementation of main concepts of & & \\ 
	    facet presentation and interaction 						& 15 &  5 \\  	\hline 	 
	
	\textbf{Visualisation (HighCharts)} &  & \\ 	\hline
	 familiarisation with the topic 								&  5 &  - \\ 	\hline 
	 adaptions, configurations 										& 15 &  - \\ 	\hline 	 
	
\end{tabular} 
\caption{Table of tasks with time and effort}
\label{tab:efforts}
\end{table}

As can be seen in table \ref{tab:efforts}, the two use cases had some common aspects in terms of architecture, module implementation, and interfaces and therefore greater parts of the implementation just had to be adapted. Above that the second use case had a simpler structure and we could control all processing steps.

The following implementations posed the largest efforts in the first use case:
\begin{itemize}
	\item the concept of Solr sub-objects and appropriate requests, e.g., to query the cardinality of resulting facets;
	\item the implementation of complex facets like temporal relations between laboratory values and other events or relations between endpoints;
	\item the handling of the poor data quality of the \tbase\ which resulted from free text input and mixed data types (done mostly within the indexing procedures);
	\item the construction of models and the processing of data requested from Solr to build various user interface constructs.
\end{itemize}

In the second use case the most time-consuming tasks were:
\begin{itemize}
	\item implementation of a local UIMA pipeline from scratch, without external analysis engines (including rebuilding the syntactic analysis); 
	\item implementation of further engines to extract additional types of annotations (new dictionaries);
	\item adaptation of result presentation (patients with multiple finding documents).
\end{itemize}

Expected adaptation costs for further use cases mainly depend on whether there are new types of annotations, new types of analysis engines (pattern-based, rule-based, machine-learning-based etc.), and other complex facet creation or visualisation requirements.

\section{Conclusion and Future Work}

We proposed a new integrated decision support system based on textual information extraction, faceted search, and information visualisation.
We tested the system on our two use cases of transplant medicine in nephrology and patient data of mammography findings. Based on freely-available open-source software tools and exchangeable information extraction modules, (two versions of) a suitable decision-support tool for the doctor has been created: this type of a knowledge based system provides physicians with a practicable tool for faceted search and hence for the analysis of medical data and decision support for cohort selection. We developed a user interface for faceted search which is based on the Solr Engine and UIMA.  We provided extensive information about the crucial aspect of implementing the architecture and visualising search results in this interactive application: the use case of nephrology showed a web-based interaction-based decision support by integrating existing structural information about patients and treatments which include numerical values, in relation to laboratory values and medications. The use case of mammography featured an adapted faceted search application  on results of an adapted information extraction pipeline. 

We conclude that the modular structure of used open-source tools software modules fits very well different domain requirements, with reasonable amount of adaptations for new use cases:
\begin{itemize}
	\item medical IE modules are rather domain specific, new modules are necessary for other text types (but UIMA is very flexible, AEs easy to exchange or add). 
	\item Solr is flexible and provides index procedures that can be adapted to different sources, based on a proper declaration of appropriate data base fields (schemata). 
	\item Faceted Search: adaption to fields is possible as facets are relatively easy to configure, but the presentation of search results must be adapted, too,  and this poses a considerable overhead. 
\end{itemize}
In general, it can be said that  the proposed  combination of modules can help physicians in various aspects of decision support by advancing information retrieval  with information extraction and intuitive user interface capabilities. 

Additional case studies will be necessary to characterise the effectiveness in clinical use. We recognise that there are limitations to our approach, e.g., the need for more user control of the information extraction process and the quality checks of the results when many patient files are processed at the same time (100,000 in the mammography use case). On the other hand, the new possibilities to include real-time information extraction results and faceted search opens many doors for future research. We are planning a targeted ANOVA based and controlled study. We are also starting to work on additional case studies in a third use case: digital mammography \cite{Sonntag2015, PrangeSS17} with medical pictures, sketches, image labels, controlled text entries, and free text entries, all based on further preprocessing steps such as hand-writing recognition. The mechanism of dealing with user feedback on the results of the information extraction process is to be extended towards interactive machine learning in the medical domain in future versions.

\section*{Acknowledgements}
This research is part of the project ``clinical data intelligence'' (KDI) which is funded by the Federal Ministry for Economic Affairs and Energy (BMWi). 

\bibliography{AIMed}

\begin{thebibliography}{10}
\expandafter\ifx\csname url\endcsname\relax
  \def\url#1{\texttt{#1}}\fi
\expandafter\ifx\csname urlprefix\endcsname\relax\def\urlprefix{URL }\fi
\expandafter\ifx\csname href\endcsname\relax
  \def\href#1#2{#2} \def\path#1{#1}\fi

\bibitem{SchmidtPS16}
D.~Schmidt, H.~Profitlich, D.~Sonntag,
  \href{http://ceur-ws.org/Vol-1613/paper_7.pdf}{Towards integrated information
  extraction and facetted search applications in nephrology}, in: Joint
  Proceedings of the 2th Workshop on Emotions, Modality, Sentiment Analysis and
  the Semantic Web and the 1st International Workshop on Extraction and
  Processing of Rich Semantics from Medical Texts co-located with {ESWC} 2016,
  Heraklion, Greece, May 29, 2016., 2016.
\newline\urlprefix\url{http://ceur-ws.org/Vol-1613/paper_7.pdf}

\bibitem{DBLP:conf/cbms/SonntagP17}
D.~Sonntag, H.~Profitlich,
  \href{https://doi.org/10.1109/CBMS.2017.119}{Integrated decision support by
  combining textual information extraction, facetted search and information
  visualisation}, in: Bamidis et~al.  \cite{DBLP:conf/cbms/2017}, pp. 95--100.
\newblock \href {http://dx.doi.org/10.1109/CBMS.2017.119}
  {\path{doi:10.1109/CBMS.2017.119}}.
\newline\urlprefix\url{https://doi.org/10.1109/CBMS.2017.119}

\bibitem{ad15}
P.~Odom, V.~Bangera, T.~Khot, D.~Page, S.~Natarajan, Extracting adverse drug
  events from text using human advice, in: Artificial Intelligence in Medicine
  - 15th Conference on Artificial Intelligence in Medicine, {AIME} 2015, Pavia,
  Italy, June 17-20, 2015. Proceedings, 2015, pp. 195--204.

\bibitem{sym15}
J.~M{\'{e}}tivier, L.~Serrano, T.~Charnois, B.~Cuissart, A.~Widl{\"{o}}cher,
  Automatic symptom extraction from texts to enhance knowledge discovery on
  rare diseases, in: Artificial Intelligence in Medicine - 15th Conference on
  Artificial Intelligence in Medicine, {AIME} 2015, Pavia, Italy, June 17-20,
  2015. Proceedings, 2015, pp. 249--254.

\bibitem{sonntag03}
S.~Vintar, L.~Todorovski, D.~Sonntag, P.~Buitelaar, Evaluating context features
  for medical relation mining, in: Proceedings of the ECML/PKDD Workshop on
  Data Mining and Text Mining for Bioinformatics, 2003.

\bibitem{MS14}
T.~Mkrtchyan, D.~Sonntag, Deep parsing at the {CLEF2014} {IE} task, in: Working
  Notes for {CLEF} 2014 Conference, Sheffield, UK, September 15-18, 2014.,
  2014, pp. 138--146.

\bibitem{SonntagWBZ09}
D.~Sonntag, P.~Wennerberg, P.~Buitelaar, S.~Zillner, Pillars of ontology
  treatment in the medical domain, J. Cases on Inf. Techn. 11~(4) (2009)
  47--73.

\bibitem{Alicante2016}
A.~Alicante, Unsupervised entity and relation extraction from clinical records
  in italian, Computers in Biology and Medicine 72~(1) (2016) 263--275.

\bibitem{DBLP:journals/nle/KlueglTBFP16}
P.~Kluegl, M.~Toepfer, P.~Beck, G.~Fette, F.~Puppe,
  \href{https://doi.org/10.1017/S1351324914000114}{{UIMA} ruta: Rapid
  development of rule-based information extraction applications}, Natural
  Language Engineering 22~(1) (2016) 1--40.
\newblock \href {http://dx.doi.org/10.1017/S1351324914000114}
  {\path{doi:10.1017/S1351324914000114}}.
\newline\urlprefix\url{https://doi.org/10.1017/S1351324914000114}

\bibitem{DBLP:conf/cbms/BretschneiderZH17}
C.~Bretschneider, S.~Zillner, M.~Hammon, P.~Gass, D.~Sonntag,
  \href{https://doi.org/10.1109/CBMS.2017.138}{Automatic extraction of breast
  cancer information from clinical reports}, in: Bamidis et~al.
  \cite{DBLP:conf/cbms/2017}, pp. 213--218.
\newblock \href {http://dx.doi.org/10.1109/CBMS.2017.138}
  {\path{doi:10.1109/CBMS.2017.138}}.
\newline\urlprefix\url{https://doi.org/10.1109/CBMS.2017.138}

\bibitem{Yee:2003}
K.-P. Yee, K.~Swearingen, K.~Li, M.~Hearst,
  \href{http://doi.acm.org/10.1145/642611.642681}{Faceted metadata for image
  search and browsing}, in: Proceedings of the SIGCHI Conference on Human
  Factors in Computing Systems, CHI '03, ACM, New York, NY, USA, 2003, pp.
  401--408.
\newblock \href {http://dx.doi.org/10.1145/642611.642681}
  {\path{doi:10.1145/642611.642681}}.
\newline\urlprefix\url{http://doi.acm.org/10.1145/642611.642681}

\bibitem{Ben-Yitzhak:2008}
O.~Ben-Yitzhak, N.~Golbandi, N.~Har'El, R.~Lempel, A.~Neumann, S.~Ofek-Koifman,
  D.~Sheinwald, E.~Shekita, B.~Sznajder, S.~Yogev,
  \href{http://doi.acm.org/10.1145/1341531.1341539}{Beyond basic faceted
  search}, in: Proceedings of the 2008 International Conference on Web Search
  and Data Mining, WSDM '08, ACM, New York, NY, USA, 2008, pp. 33--44.
\newblock \href {http://dx.doi.org/10.1145/1341531.1341539}
  {\path{doi:10.1145/1341531.1341539}}.
\newline\urlprefix\url{http://doi.acm.org/10.1145/1341531.1341539}

\bibitem{Stoica2007Automating}
E.~Stoica, M.~Hearst, M.~Richardson, Automating creation of hierarchical
  faceted metadata structures, in: NAACL/HLT 2007, 2006.

\bibitem{HearstFacettedBrowsing}
M.~A. Hearst, Design recommendations for hierarchical faceted search
  interfaces, in: Proc. SIGIR 2006, Workshop on Faceted Search, 2006, pp.
  26--30.

\bibitem{sacco2006}
G.~Sacco, Dynamic taxonomies and guided searches, Journal of the American
  Society for Information Science and Technology 57~(6) (2006) 792--796.

\bibitem{sacco2015}
G.~Sacco, Dynamic taxonomies for intelligent information access, in:
  M.~Khosrow-Pour (Ed.), Encyclopedia of Information Science and Technology,
  3rd Edition, 2014, pp. 3883--3892.

\bibitem{Wongsuphasawat:2011}
K.~Wongsuphasawat, J.~A. Guerra~G\'{o}mez, C.~Plaisant, T.~D. Wang,
  M.~Taieb-Maimon, B.~Shneiderman,
  \href{http://doi.acm.org/10.1145/1978942.1979196}{Lifeflow: Visualizing an
  overview of event sequences}, in: Proceedings of the SIGCHI Conference on
  Human Factors in Computing Systems, CHI '11, ACM, New York, NY, USA, 2011,
  pp. 1747--1756.
\newblock \href {http://dx.doi.org/10.1145/1978942.1979196}
  {\path{doi:10.1145/1978942.1979196}}.
\newline\urlprefix\url{http://doi.acm.org/10.1145/1978942.1979196}

\bibitem{EventFlow16}
F.~Du, B.~Shneiderman, C.~Plaisant, S.~Malik, A.~Perer, Coping with volume and
  variety in temporal event sequences: Strategies for sharpening analytic
  focus, IEEE Transactions on Visualization and Computer Graphics PP~(99)
  (2016) 1--1.
\newblock \href {http://dx.doi.org/10.1109/TVCG.2016.2539960}
  {\path{doi:10.1109/TVCG.2016.2539960}}.

\bibitem{Monroe:2013}
M.~Monroe, R.~Lan, J.~Morales~del Olmo, B.~Shneiderman, C.~Plaisant,
  J.~Millstein, \href{http://doi.acm.org/10.1145/2470654.2481325}{The
  challenges of specifying intervals and absences in temporal queries: A
  graphical language approach}, in: Proceedings of the SIGCHI Conference on
  Human Factors in Computing Systems, CHI '13, ACM, New York, NY, USA, 2013,
  pp. 2349--2358.
\newblock \href {http://dx.doi.org/10.1145/2470654.2481325}
  {\path{doi:10.1145/2470654.2481325}}.
\newline\urlprefix\url{http://doi.acm.org/10.1145/2470654.2481325}

\bibitem{Malik:2016}
S.~Malik, B.~Shneiderman, F.~Du, C.~Plaisant, M.~Bjarnadottir,
  \href{http://doi.acm.org/10.1145/2890478}{High-volume hypothesis testing:
  Systematic exploration of event sequence comparisons}, ACM Trans. Interact.
  Intell. Syst. 6~(1) (2016) 9:1--9:23.
\newblock \href {http://dx.doi.org/10.1145/2890478}
  {\path{doi:10.1145/2890478}}.
\newline\urlprefix\url{http://doi.acm.org/10.1145/2890478}

\bibitem{Malik:2015}
S.~Malik, F.~Du, M.~Monroe, E.~Onukwugha, C.~Plaisant, B.~Shneiderman,
  \href{http://doi.acm.org/10.1145/2678025.2701407}{Cohort comparison of event
  sequences with balanced integration of visual analytics and statistics}, in:
  Proceedings of the 20th International Conference on Intelligent User
  Interfaces, IUI '15, ACM, New York, NY, USA, 2015, pp. 38--49.
\newblock \href {http://dx.doi.org/10.1145/2678025.2701407}
  {\path{doi:10.1145/2678025.2701407}}.
\newline\urlprefix\url{http://doi.acm.org/10.1145/2678025.2701407}

\bibitem{schroeterTBase}
K.~Schr\"{o}ter, Tbase2, a web-based electronic patient record., Fundamenta
  Informaticae 43~(1-4) (2000) 343--353.

\bibitem{lindemannTBase}
G.~Lindemann, A web-based patient record for hospitals - the design of tbase2,
  in: H.-P. Bruch (Ed.), New Aspects of Hight Technology in Medicine: Hannover
  (Germany), Monduzzi Editore, International Proceedings Division, 2000, pp.
  409--414.

\bibitem{OASIS:UIMA:2009}
D.~Ferrucci, A.~Lally, K.~Verspoor, E.~Nyberg,
  \href{https://docs.oasis-open.org/uima/v1.0/uima-v1.0.html}{Unstructured
  information management architecture ({UIMA}) version 1.0}, OASIS Standard
  (mar 2009).
\newline\urlprefix\url{https://docs.oasis-open.org/uima/v1.0/uima-v1.0.html}

\bibitem{DBLP:conf/clef/MkrtchyanS14}
T.~Mkrtchyan, D.~Sonntag,
  \href{http://ceur-ws.org/Vol-1180/CLEF2014wn-eHealth-MkrtchyantEt2014.pdf}{Deep
  parsing at the {CLEF2014} {IE} task}, in: L.~Cappellato, N.~Ferro, M.~Halvey,
  W.~Kraaij (Eds.), Working Notes for {CLEF} 2014 Conference, Sheffield, UK,
  September 15-18, 2014., Vol. 1180 of {CEUR} Workshop Proceedings,
  CEUR-WS.org, 2014, pp. 138--146.
\newline\urlprefix\url{http://ceur-ws.org/Vol-1180/CLEF2014wn-eHealth-MkrtchyantEt2014.pdf}

\bibitem{UIMABook}
\href{https://uima.apache.org/d/ruta-2.6.1/tools.ruta.book.html\#ugr.tools.ruta.integration}{Apache
  uima ruta guide and reference} (2017).
\newline\urlprefix\url{https://uima.apache.org/d/ruta-2.6.1/tools.ruta.book.html\#ugr.tools.ruta.integration}

\bibitem{Sonntag2015}
D.~Sonntag, V.~Tresp, S.~Zillner, A.~Cavallaro, M.~Hammon, A.~Reis, A.~P.
  Fasching, M.~Sedlmayr, T.~Ganslandt, H.-U. Prokosch, K.~Budde, D.~Schmidt,
  C.~Hinrichs, T.~Wittenberg, P.~Daumke, G.~P. Oppelt, The clinical data
  intelligence project, Informatik-Spektrum Journal (2015) 1--11.

\bibitem{PrangeSS17}
A.~Prange, D.~Schmidt, D.~Sonntag,
  \href{https://doi.org/10.1109/CBMS.2017.132}{A digital pen based tool for
  instant digitisation and digitalisation of biopsy protocols}, in: Bamidis
  et~al.  \cite{DBLP:conf/cbms/2017}, pp. 773--774.
\newblock \href {http://dx.doi.org/10.1109/CBMS.2017.132}
  {\path{doi:10.1109/CBMS.2017.132}}.
\newline\urlprefix\url{https://doi.org/10.1109/CBMS.2017.132}

\bibitem{DBLP:conf/cbms/2017}
P.~D. Bamidis, S.~T. Konstantinidis, P.~P. Rodrigues (Eds.),
  \href{http://ieeexplore.ieee.org/xpl/mostRecentIssue.jsp?punumber=8100282}{30th
  {IEEE} International Symposium on Computer-Based Medical Systems, {CBMS}
  2017, Thessaloniki, Greece, June 22-24, 2017}, {IEEE} Computer Society, 2017.
\newline\urlprefix\url{http://ieeexplore.ieee.org/xpl/mostRecentIssue.jsp?punumber=8100282}

\end{thebibliography}

\end{document}